\documentclass[aps,twocolumn,nofootinbib,showpacs]{revtex4}
\usepackage{bm,amssymb,epsf}
\newcommand{\kB}{k_{\rm B}}

\newcommand{\ave}[1]{\left\langle#1\right\rangle_\rho}
\newcommand{\aveb}[1]{\left\langle#1\right\rangle}
\newcommand{\Dbra}[1]{\left\langle#1\right|}
\newcommand{\Dket}[1]{\left|#1\right\rangle}
\newcommand{\cancor}[2]{\left\langle#1;#2\right\rangle_\rho}
\newcommand{\Qcommu}[2]{[#1,#2]}

\newcommand{\Qantico}[2]{\{#1,#2\}}
\newcommand{\CPoiss}[2]{\bm{\{}#1,#2\bm{\}}}
\newcommand{\Cdissip}[2]{\mbox{$\bm{[\hspace{-0.28em}[}$}#1,#2\mbox{$\bm{]\hspace{-0.28em}]}$}}

\begin{document}
\bibliographystyle{apsrev}

\title{Nonlinear thermodynamic quantum master equation: Properties and examples}

\author{Hans Christian \"Ottinger}
\email[]{hco@mat.ethz.ch}
\homepage[]{http://www.polyphys.mat.ethz.ch/}
\affiliation{ETH Z\"urich, Department of Materials, Polymer Physics, HCI H 543,
CH-8093 Z\"urich, Switzerland}

\date{\today}

\begin{abstract}
The quantum master equation obtained from two different thermodynamic arguments is seriously nonlinear. We argue that, for quantum systems, nonlinearity occurs naturally in the step from reversible to irreversible equations and we analyze the nature and consequences of the nonlinear contribution. The thermodynamic nonlinearity naturally leads to canonical equilibrium solutions and extends the range of validity to lower temperatures. We discuss the Markovian character of the thermodynamic quantum master equation and introduce a solution strategy based on coupled evolution equations for the eigenstates and eigenvalues of the density matrix. The general ideas are illustrated for the two-level system and for the damped harmonic oscillator. Several conceptual implications of the nonlinearity of the thermodynamic quantum master equation are pointed out, including the absence of a Heisenberg picture and the resulting difficulties with defining multi-time correlations.
\end{abstract}




\maketitle

\section{Introduction}
Quantum master equations provide a useful tool for describing dissipative quantum systems \cite{BreuerPetru,Weiss}. Most popular are the linear master equations of the Lindblad form \cite{Lindblad76}. However, it has been known for some 30 years that these equations have a fundamental problem because they invoke an incorrect ``quantum regression hypothesis'' \cite{Grabert82,GraberTalkner83,Talkner86}. For quantum systems in contact with a heat bath, this problem has been overcome by introducing a nonlinear master equation associated with a ``modified quantum regression hypothesis'' \cite{Grabert82}. This modified master equation has been obtained by means of the projection-operator method. The resulting nonlinear master equation is not limited to high temperatures for which quantum effects are unavoidably small. It has actually been shown that the quantum master equation can be applied down to arbitrarily low temperatures provided that the frictional coupling to the heat bath becomes sufficiently weak \cite{Grabert06}. Once the thermodynamically consistent nonlinear master equation has been formulated, one can look for special situations in which exact or approximate linear master equations can be derived. This has been done in \cite{KarrleinGrabert97}, with the conclusion that the resulting master equations are not of the popular Lindblad form.

The formulation of a nonlinear master equation in \cite{Grabert82} was triggered by a problem with the ``quantum regression hypothesis,'' that is, by a thermodynamic argument. The field of quantum dissipation has recently been approached from an entirely different perspective, which is also rooted in thermodynamics and leads to a more general nonlinear quantum master equation \cite{hco199}. Starting from a modern geometric formulation of nonequilibrium thermodynamics for classical systems \cite{hco99,hco100,hcobet}, Dirac's method of classical analogy (see Chapter~IV of \cite{Dirac}) has been employed for a generalization to quantum systems. For reversible systems, the recognition of the deep correspondence between classical Poisson brackets and quantum commutators is the key to establishing the quantum-classical correspondence. Poisson brackets provide one of the important structures used in nonequilibrium thermodynamics, namely to formulate reversible dynamics. Irreversible dynamics is formulated in terms of dissipative brackets, for which a quantum generalization in terms of canonical correlations has been proposed in \cite{hco199}. The resulting evolution equations may be considered as a generalization of the nonlinear quantum master equation proposed in \cite{Grabert82}: rather than being restricted to heat baths, the formulation of \cite{hco199} is applicable to arbitrary thermodynamic environments, including time-dependent ones, where also the influence of the quantum subsystem on the evolution of the classical environment is predicted.

Whereas linearity seems natural in a quantum mechanical setting, it should not be taken for granted in thermodynamics. This is a consequence of the appearance of entropy, which typically involves logarithmic terms. Going beyond reversibility in general requires to go beyond linearity. For classical systems, there occurs a fortuitous cancelation that leads to the linearity of the Fokker-Planck equation (on the level of distribution functions $f$, the key identity is $f d (\delta S_f/\delta f) = - \kB f d \ln f = - \kB d f$, where $S_f$ is the entropy and $\kB$ is Boltzmann's constant; see Section III.B of \cite{hco99}). The noncommutativity of quantum observables prevents such a cancelation. As thermodynamics is the language for formulating healthy equations with well-behaved solutions, the nonlinearity, which we have recognized as a quantum effect required by the principles of thermodynamics, should not at all be considered as a drawback. The purpose of the present paper is to elaborate the advantages of the nonlinear thermodynamic master equation originally proposed in \cite{Grabert82} and recently recovered as a special case of \cite{hco199} for a heat bath in a detailed comparison with the popular linear master equation. We consider the two-level system and the damped quantum harmonic oscillator as concrete examples.

We first summarize the equations obtained from thermodynamics for quantum systems interacting with a classical environment and discuss some of their key features, most importantly, nonlinearity (Sec.~\ref{sec2}). After a short description of possible solution strategies for the nonlinear master equations resulting from thermodynamics (Sec.~\ref{sec3}), we study the examples of the relaxation behavior of the two-level system (Sec.~\ref{sec4p}) and the damped harmonic oscillator (Sec.~\ref{sec4}) in detail. We end with a brief summary and offer some concluding remarks (Sec.~\ref{sec5}).

\section{Thermodynamic approach to quantum dissipation}\label{sec2}
As a first step, we discuss different motivations for considering quantum-classical systems. We then discuss the ``quantum regression hypothesis'' and compile the essential results of the thermodynamic approach to quantum dissipation. In particular, we give the evolution equations for the quantum subsystem and its environment, and we discuss some basic features of these equations.

\begin{figure}
\centerline{\epsfxsize=8.6cm \epsffile{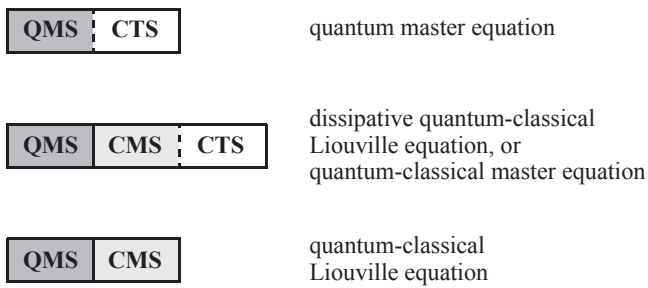}}
\caption[ ]{Different couplings between quantum mechanical systems (QMS), classical mechanical systems (CMS), and classical thermodynamic systems (CTS); continuous lines between systems indicate a reversible coupling, whereas dashed lines represent an irreversible coupling.}
\label{fig_coupled_systems}
\end{figure}

\subsection{Quantum-classical systems}
There exist different reasons for coupling quantum and classical systems. We begin with the discussion of the two extremes illustrated in the bottom and top of Fig.~\ref{fig_coupled_systems}: (i) The reversible coupling of a quantum system to classical phase space (``atomistic'') variables, and (ii) the irreversible coupling of a quantum system to a classical thermodynamic system, usually a heat bath. The motivation behind the approach (i) is to make ``atomistic'' simulations feasible by treating some of their degrees of freedom classically, typically because they are associated with the heavier particles in the system. The combined system (i) is described by a purely reversible \emph{quantum-classical Liouville equation} (see, for example, Eq.~(6) of \cite{Kapral06}). The approach (ii) is based on the elimination of degrees of freedom from the environment, which typically requires a weak coupling so that the quantum subsystem feels the influence of the environment only on long time scales and rapidly fluctuating variables can hence be eliminated from the environment. Of course, some kind of approximation or limit is always needed to justify the treatment of certain degrees of freedom by classical mechanics or thermodynamics. Typical applications are quantum rate processes (such as proton or electron transport) depending on the presence of characteristic groups of atoms for (i), and nuclear spin relaxation in a surrounding medium for (ii).

Quantum-classical Liouville equations are usually simplified by means of the momentum-jump approximation for nonadiabatic transitions and the corresponding changes of the bath momentum \cite{Kapral06}. For the proper description of decoherence, it is necessary to include further classical degrees of freedom which typically serve as a heat bath for the classical atoms as illustrated in the middle of Fig.~\ref{fig_coupled_systems}. One then arrives at the \emph{dissipative quantum-classical Liouville equation} involving a Fokker-Planck-like operator (see, for example, Eq.~(15) of \cite{Kapral06}). The irreversible coupling between the two classical subsystems may contain an additional reversible contribution, such as an effective force. As an alternative to the dissipative quantum-classical Liouville equation, a Markovian quantum-classical master equation for the three coupled systems in the middle of Fig.~\ref{fig_coupled_systems} has been derived on the full phase space, including a typical molecule of a heat bath as the classical thermodynamic system (see Eq.~(27) of \cite{GrunwaldKapral07} and Eq.~(20) of \cite{RankKapral10}).

In the present work, we are interested in the approach (ii), that is, in the irreversible coupling of a quantum system to a classical thermodynamic environment. As discussed in the introduction, thermodynamic consistency arguments have been discussed and implemented in the context of the \emph{quantum master equation}. Also for other approaches to dissipative quantum systems, such as operator Langevin equations, stochastic dynamics in Hilbert space, or path integrals \cite{Weiss,BreuerPetru,HanggiIngold05,FeynmanVernon63,Grabertetal88}, thermodynamic consistency should be established.

The nonlinear thermodynamic master equation of \cite{hco199} describes a quantum system interacting with a classical thermodynamic system where, in general, the classical environment is a nonequilibrium system with its own thermodynamic evolution that has to be determined together with the evolution of the quantum system. The coupling between the quantum subsystem and its classical environment is entirely of the irreversible type; the postulated Markovian description implies restrictions on the separation of time scales and the weakness of the interaction \cite{Grabert06}. For the special case of an equilibrium environment, that is, a heat bath with a fixed temperature, the nonlinear thermodynamic master equation has been derived already 30 years ago by means of the projection-operator method \cite{Grabert82}. As thermodynamically consistent master equations have hardly ever been used in the literature, our goal is to consider the properties of these nonlinear master equations in some detail.

\subsection{Quantum regression hypothesis}\label{sec2B0}
Quantum master equations for the evolution of the density matrix or statistical operator $\rho$ on a suitable Hilbert space are usually assumed to be of the linear form \cite{BreuerPetru,Weiss},
\begin{equation}\label{linearme}
    \frac{d\rho}{dt} = - i {\cal L} \rho ,
\end{equation}
where $\cal L$ is a suitable super-operator, say of the Lindblad form \cite{Lindblad76}. According to the Schr\"odinger picture, the time-dependent density matrix $\rho$ can be used to calculate the evolving average $\ave{A} = {\rm tr}(A\rho)$ of an observable that is represented by a time-independent self-adjoint operator $A$. The next goal is to calculate multi-time correlation functions. In order to do so, one usually switches to the Heisenberg picture based on evolving observables to be averaged with a time-independent density matrix. When two-time correlations are evaluated by means of the Heisenberg picture, one finds that the decay of two-time correlations is governed by exactly the same evolution equation as the decay of averages, which is known as the ``quantum regression hypothesis'' (see, for example, Section 3.2.4 of \cite{BreuerPetru}). More precisely, we have for example the following two-time correlation of two observables $A$ and $B$,
\begin{equation}\label{Heispiccomm}
    \ave{\Qcommu{A(t)}{B}} = {\rm tr} \left( A e^{-i{\cal L}t} \Qcommu{B}{\rho} \right) ,
\end{equation}
which nicely shows the occurrence of the evolution super-operator in this expression for two-time correlations.

An alternative possibility to calculate two-time correlations is based on the fluctuation-dissipation theorem of the first kind (see Eq.~(4.2.18) of \cite{KuboetalII} or Eqs.~(6.7), (6.11), and (6.14) of \cite{Grabert82}),
\begin{equation}\label{FDTcomm}
    \ave{\Qcommu{A(t)}{B}} = \frac{\hbar}{\kB T_{\rm e}} \,
    {\rm tr} \left( A e^{-i{\cal L}t} {\cal L} B_\rho \right) ,
\end{equation}
where $\hbar$ and $\kB$ are Planck's constant divided by $2\pi$ and Boltzmann's constant, respectively, and $T_{\rm e}$ is the temperature parameter of the canonical density matrix $\rho$ assumed in Eq.~(\ref{FDTcomm}). For given operator $A$ and density matrix $\rho$, the operator $A_\rho$ is basically the product of $A$ and $\rho$, but with a compromise between placing $\rho$ to the left or the right of $A$,
\begin{equation}\label{Atildef}
    A_\rho = \int_0^1  \rho^\lambda A \, \rho^{1-\lambda} \, d\lambda .
\end{equation}
If $A$ is self-adjoint, this property is inherited by $A_\rho$. Note that $A_\rho$ has the useful property
\begin{equation}\label{lnLemma1}
     \Qcommu{A}{\rho} = \Qcommu{A_\rho}{\ln\rho} ,
\end{equation}
which follows from looking at arbitrary matrix elements formed with the eigenstates of the density matrix and performing the elementary integration over $\lambda$ in Eq.~(\ref{Atildef}).

The fluctuation-dissipation theorem can be derived by multiplying Eq.~(\ref{lnLemma1}) with another observable and taking the trace,
\begin{equation}\label{lnLemma2}
    \ave{\Qcommu{A}{B}} = {\rm tr} \left( A \Qcommu{B_\rho}{\ln\rho} \right) .
\end{equation}
If $\rho$ is a canonical density matrix, $H = - \kB T_{\rm e} \ln \rho$ describes the Hamiltonian time evolution, and $A$ is the Heisenberg operator $A(t)$, then Eq.~(\ref{lnLemma2}) yields the fluctuation-dissipation theorem (\ref{FDTcomm}). If the operator identity (\ref{lnLemma1}) is used in Eq.~(\ref{Heispiccomm}) one realizes that the ``quantum regression hypothesis'' can only be consistent with the fluctuation-dissipation theorem (\ref{FDTcomm}) if $\cal L$ is of the Hamiltonian form, but not for dissipative master equations. This observation is the well known failure of the ``quantum regression hypothesis.''

This failure of the ``quantum regression hypothesis'' and the nonlinear dependence of $B_\rho$ on $\rho$ in Eq.~(\ref{FDTcomm}) motivated Grabert to revisit the standard projection-operator derivation of quantum master equations \cite{Grabert} with a relevant density matrix of the exponential form, where the deviation from the Hamiltonian in the exponent can be interpreted as the thermodynamic force operator conjugate to the density matrix. In the Markovian limit, the resulting equation is of the nonlinear form (see Eq.~(5.22) of \cite{Grabert82}),
\begin{equation}\label{Grabertme}
    \frac{d\rho}{dt} = \frac{i}{\hbar} \Qcommu{\rho}{H}
    - \frac{M}{\kB T_{\rm e}} \, \Qcommu{Q}{\Qcommu{Q}{H}_\rho}
    - M \, \Qcommu{Q}{\Qcommu{Q}{\rho}} ,
\end{equation}
with a suitable parameter $M$ describing the strength of the dissipation and an observable $Q$ describing the interaction between the quantum subsystem and its quantum environment. Note that the temperature $T_{\rm e}$ is the only parameter characterizing the state of the environment, which hence acts as a heat bath. The strength of the dissipative interaction between the subsystem and its environment is characterized by the parameter $M$. Equation (\ref{Grabertme}) may be addressed as a thermodynamic master equation because it has been derived with a relevant density matrix characterized in terms of a thermodynamic force operator and because, as a consequence, it is consistent with the fluctuation-dissipation theorem.

\subsection{Thermodynamic quantum master equation}\label{sec2B}
The thermodynamic quantum master equation (\ref{Grabertme}) holds for an environment acting as a heat bath. This nonlinear master equation can be generalized to more complicated classical nonequilibrium systems as environments. Based on purely thermodynamic considerations and a generalization from classical to quantum systems inspired by a geometric formulation of nonequilibrium thermodynamics, the following master equation for the evolution of the density matrix or statistical operator $\rho$ has been proposed to characterize a quantum subsystem in contact with an arbitrary classical nonequilibrium system acting as its environment:
\begin{eqnarray}
    \frac{d\rho}{dt} = \frac{i}{\hbar} \Qcommu{\rho}{H}
    &-& \frac{1}{\kB} \Cdissip{H_{\rm e}}{S_{\rm e}}^Q_x \, \Qcommu{Q}{\Qcommu{Q}{H}_\rho}
    \nonumber\\
    &-& \Cdissip{H_{\rm e}}{H_{\rm e}}^Q_x \, \Qcommu{Q}{\Qcommu{Q}{\rho}} .
    \qquad
\label{GENERICme}
\end{eqnarray}
The first term describes the reversible contribution to the evolution generated by the Hamiltonian $H$ via the commutator. All other terms are of irreversible nature and result from a coupling of the quantum subsystem to its environment. They are expressed through double commutators involving the self-adjoint coupling operator $Q$ so that the normalization condition, ${\rm tr}\,\rho=1$, is automatically preserved in time. As a consequence of the occurrence of commutators with the coupling operator $Q$, the evolution of the average $\ave{Q}$ performed with the time-dependent density matrix $\rho$  is not explicitly affected by the dissipative terms.

Whereas the type of the coupling is given by the observable $Q$, the strength of the coupling is expressed in a dissipative bracket $\Cdissip{\,}{\,}$ defined as a binary operation on the space of observables for the classical environment (throughout this work, boldface bracket symbols are used to distinguish classical dissipative and Poisson brackets from quantum commutators and anticommutators, respectively). If the equilibrium or nonequilibrium states of the environment are characterized by state variables $x$, classical observables are functions or functionals of $x$, and their evaluation at a particular point of the state space is indicated by the subscript $x$. The classical observables $H_{\rm e}$ and $S_{\rm e}$ in Eq.~(\ref{GENERICme}) are the energy and the entropy of the environment, respectively. Dissipative brackets are commonly used to characterize the entropy production rate in nonequilibrium thermodynamics \cite{BerisEdwards,hcobet}. They are characterized by the following properties: $\Cdissip{A_{\rm e}}{B_{\rm e}}$ is bilinear in $A_{\rm e}$ and $B_{\rm e}$, is symmetric,
\begin{equation}\label{dissbrasym}
    \Cdissip{A_{\rm e}}{B_{\rm e}} = \Cdissip{B_{\rm e}}{A_{\rm e}} ,
\end{equation}
as well as nonnegative,
\begin{equation}\label{dissbrapos}
    \Cdissip{A_{\rm e}}{A_{\rm e}} \geq 0 ,
\end{equation}
and satisfies the Leibniz or product rule,
\begin{equation}\label{dissbraLeib}
    \Cdissip{A_{\rm e} B_{\rm e}}{C_{\rm e}} =
    A_{\rm e} \Cdissip{B_{\rm e}}{C_{\rm e}}
    + B_{\rm e} \Cdissip{A_{\rm e}}{C_{\rm e}} ,
\end{equation}
for arbitrary environmental variables $A_{\rm e}$, $B_{\rm e}$, and $C_{\rm e}$. As a straightforward generalization of Eq.~(\ref{GENERICme}), several coupling operators $Q_j$ can be incorporated easily \cite{hco199,Grabert82}.

\subsection{Nonlinearity}
A most striking feature of the thermodynamic quantum master equation is its nonlinearity in $\rho$. In view of the definition (\ref{Atildef}) of $A_\rho$, the second term in Eq.~(\ref{GENERICme}) will, in general, be nonlinear in $\rho$. This definition can be rewritten in a form that brings out the relationship to another possible compromise in placing $\rho$ and extracts the nonlinearity,
\begin{equation}\label{Atildefa1}
    A_\rho = \frac{1}{2} \left( A \rho + \rho A + A'_\rho \right) ,
\end{equation}
with the nonlinear term
\begin{equation}\label{Atildefa2}
    A'_\rho =  - \int_0^1 \Qcommu{\rho^\lambda}{\Qcommu{\rho^{1-\lambda}}{A}} \, d\lambda .
\end{equation}
Setting $A'_\rho = 0$ corresponds to a linearization of the master equation (\ref{GENERICme}) which has been proposed, but not recommended, in \cite{hco199}. We hence may think of $A'_\rho$ as the origin of nonlinearity in the full thermodynamically consistent quantum master equation. The occurrence of $A_\rho$ is a consequence of employing canonical correlations $\cancor{\,}{\,}$ (see Eq.~(4.1.12) of \cite{KuboetalII}) as the key structural element in the generalization of dissipative brackets from classical to quantum systems,
\begin{equation}\label{cancor}
    \cancor{A}{B} = \int_0^1 {\rm tr}
    \big( \rho^\lambda A \, \rho^{1-\lambda} B \big) \, d\lambda
    = {\rm tr} \big( A_\rho B \big) .
\end{equation}
The canonical correlation is symmetric, $\cancor{A}{B} = \cancor{B}{A}$, and positive, $\cancor{A}{A} \geq 0$. Moreover, averages can be obtained as special cases of canonical correlations, $\ave{A} = {\rm tr}(A \rho) = {\rm tr}(A_\rho) = \cancor{A}{1}$.

The evaluation of $A_\rho$ involves the calculation of the powers $\rho^\lambda$. To handle the nonlinear quantum master equation it is hence natural to determine the eigenstates $\Dket{\pi_n}$ and the eigenvalues $p_n$ of the density matrix $\rho$ and to make use of the representation
\begin{equation}\label{densmatrix}
    \rho = \sum_n p_n \Dket{\pi_n} \Dbra{\pi_n} .
\end{equation}
In terms of the eigenstates of $\rho$, the evaluation of the matrix elements of the modified operator $A_\rho$ is straightforward because the integration over $\lambda$ can be carried out. We obtain the useful identity
\begin{equation}\label{Arhoident}
    \Dbra{\pi_m} A_\rho \Dket{\pi_n} = \frac{p_m - p_n}{\ln p_m - \ln p_n}
    \Dbra{\pi_m} A \Dket{\pi_n} ,
\end{equation}
which is equivalent to Eq.~(\ref{lnLemma1}). For the factor occurring between the matrix elements of $A_\rho$ and $A$ in Eq.~(\ref{Arhoident}), we have the inequalities
\begin{equation}\label{Arhoidentineq}
    0 \leq \frac{p_m - p_n}{\ln p_m - \ln p_n} \leq \frac{p_m+p_n}{2} \leq 1 ,
\end{equation}
where the central inequality becomes an equality if, and only if, $p_m$ and $p_n$ are equal. If we use the approximation
\begin{equation}\label{Arhoapprox}
    \frac{p_m - p_n}{\ln p_m - \ln p_n} \approx \frac{p_m+p_n}{2} ,
\end{equation}
which corresponds to setting $A'_\rho = 0$ in Eq.~(\ref{Atildefa1}), we arrive at the previously mentioned linearized master equation
\begin{eqnarray}
    \frac{d\rho}{dt} = \frac{i}{\hbar} \Qcommu{\rho}{H}
    &-& \frac{1}{2\kB} \Cdissip{H_{\rm e}}{S_{\rm e}}^Q_x \, \Qcommu{Q}{\Qantico{\Qcommu{Q}{H}}{\rho}}
    \nonumber\\
    &-& \Cdissip{H_{\rm e}}{H_{\rm e}}^Q_x \, \Qcommu{Q}{\Qcommu{Q}{\rho}} ,
\label{GENERICmelin}
\end{eqnarray}
where $\Qantico{\,}{\,}$ is the anticommutator.

The linearization obtained by turning the inequality (\ref{Arhoidentineq}) into the approximation (\ref{Arhoapprox}) may look somewhat ambiguous. It is clearly different from a systematic linearization around a given reference state, but with the attractive advantages of simplicity and generality. This linearization corresponds to replacing $A_\rho$ by $(A \rho + \rho A)/2$, which looks like a reasonable alternative to solve the problem of placing $\rho$ in the product of $A$ and $\rho$. Whereas Eq.~(\ref{GENERICmelin}) is a convenient linearization of the thermodynamic master equation (\ref{GENERICme}), we do not recommend its use because it destroys the thermodynamic structure of the original master equation. The thermodynamic structure is important for the qualitative properties of the solutions, for example, for the existence of canonical equilibrium solutions, as discussed in the following subsection.

\subsection{Heat bath}
The thermodynamic approach to quantum dissipation is valid for arbitrary environments, as long as they may be treated as classical nonequilibrium systems. We here consider the simple and important special case of a heat bath which can be described by a single independent state variable $x$, say the total energy $H_{\rm e}$. The complete thermodynamic information about this system is contained in the functional form of the entropy, $S_{\rm e}(H_{\rm e})$. In particular, we can assign a temperature $T_{\rm e}$ to the heat bath,
\begin{equation}\label{tempdef}
    \frac{1}{T_{\rm e}} = \frac{\partial S_{\rm e}(H_{\rm e})}{\partial H_{\rm e}} .
\end{equation}

The most general form of a dissipative bracket for a heat bath is given by
\begin{equation}\label{heatbathdbgen}
    \Cdissip{A_{\rm e}}{B_{\rm e}}^Q = \frac{dA_{\rm e}}{dH_{\rm e}}
    \, M(T_{\rm e}) \, \frac{dB_{\rm e}}{dH_{\rm e}} ,
\end{equation}
where $M(T_{\rm e})$ is a positive function. Such a bracket is bilinear, symmetric, positive, and satisfies the Leibniz rule, as postulated in Eqs.~(\ref{dissbrasym})--(\ref{dissbraLeib}). Any dissipative bracket for the heat bath at temperature $T_{\rm e}$ hence satisfies the additional condition
\begin{equation}\label{bathequilib}
    T_{\rm e} \Cdissip{H_{\rm e}}{S_{\rm e}}^Q =
    \Cdissip{H_{\rm e}}{H_{\rm e}}^Q ,
\end{equation}
which is exactly the condition required to obtain the canonical equilibrium solution to the quantum master equation (\ref{GENERICme}) expected for weak coupling between the quantum and classical subsystems,
\begin{equation}\label{rhoeq}
    \rho_{\rm eq} \, \propto \, \exp \left\{ - \frac{H}{\kB T_{\rm e}} \right\} .
\end{equation}
To verify this equilibrium solution, one can make use of the identity $\Qcommu{A}{\rho} = \Qcommu{A_\rho}{\ln\rho}$, which is implied by Eq.~(\ref{Arhoident}), in the last term of the master equation (\ref{GENERICme}) for $A=Q$. The guaranteed existence of the canonical equilibrium solution is a convenient advantage of the thermodynamic quantum master equation. It is deeply linked to the nonlinearity of the master equation and hence to the formulation of the dissipative bracket in terms of canonical correlations.

The general thermodynamic approach also provides an equation for the entropy production. The average $\bar{S}$ of the total entropy of the quantum subsystem and its classical environment evolves according the equation
\begin{equation}\label{entropprod1}
    \frac{d\bar{S}}{dt} = - \kB {\rm tr} \left( \ln\rho \, \frac{d\rho}{dt} \right)
    + \frac{d S_{{\rm e},x}}{dt} .
\end{equation}
For a pure heat bath at temperature $T_{\rm e}$, by construction of the dissipative bracket the entropy production can be expressed in terms of the canonical correlation \cite{hco199}
\begin{equation}\label{entropprod2}
    \frac{d\bar{S}}{dt} =  \frac{M(T_{\rm e})}{\kB T_{\rm e}^2}
    \cancor{i\Qcommu{Q}{F}}{i\Qcommu{Q}{F}} ,
\end{equation}
where
\begin{equation}\label{freeenergdef}
    F = H + \kB T_{\rm e} \ln\rho
\end{equation}
is the Helmholtz free energy operator.

\subsection{Reaction on environment}
The master equation (\ref{GENERICme}) describes the influence of a classical environment on a quantum subsystem. Of course, in response, the quantum system also has an influence on its environment. In general, the state $x$ of the environment hence varies in time and the strength of the coupling in the thermodynamic quantum master equation (\ref{GENERICme}) becomes time-dependent. By neglecting the evolution of $T_{\rm e}$ for a heat bath we implicitly assume an infinitely large heat capacity of the bath.

The thermodynamic approach actually provides a corresponding equation for the evolution of environmental observables,
\begin{eqnarray}
    \frac{dA_{{\rm e},x}}{dt} &=& \CPoiss{A_{\rm e}}{H_{\rm e}}_x
    + \Cdissip{A_{\rm e}}{S_{\rm e}}_x \nonumber\\
    &-& \frac{1}{\kB} \Cdissip{A_{\rm e}}{S_{\rm e}}^Q_x
    \cancor{\Qcommu{Q}{H}}{\Qcommu{Q}{H}} \nonumber\\
    &+& \Cdissip{A_{\rm e}}{H_{\rm e}}^Q_x \ave{\Qcommu{Q}{\Qcommu{Q}{H}}} .
\label{GENERICcla}
\end{eqnarray}
In this equation, $\CPoiss{\,}{\,}$ and $\Cdissip{\,}{\,}$ are the Poisson and dissipative brackets of the classical system, respectively \cite{hco99,hco100,hcobet}. In addition to the properties (\ref{dissbrasym})--(\ref{dissbraLeib}) of dissipative brackets, energy conservation in the environment (except for the balanced exchange of energy with the quantum subsystem) is guaranteed by the degeneracy requirement $\Cdissip{A_{\rm e}}{H_{\rm e}} = 0$ for all classical observables $A_{\rm e}$. The Poisson bracket $\CPoiss{\,}{\,}$ is bilinear, antisymmetric, and satisfies the Leibniz rule as well as the Jacobi identity, where the latter expresses the time-structure invariance of the Poisson bracket \cite{Marsden,MarsdenRatiu,hcobet,hco102}. All these properties of classical Poisson brackets are shared by their famous quantum counterparts, the commutators \cite{Dirac}.

If one looks only at the master equation (\ref{GENERICme}), the occurrence of time-dependent coefficients suggests non-Markovian behavior. If one considers the coupled evolution of the quantum subsystem and the classical environment according to the thermodynamically coupled set of Eqs.~(\ref{GENERICme}) and (\ref{GENERICcla}), however, the Markovian character of the description is restored, provided that the total system is closed. By comparing Eqs.~(\ref{GENERICme}) and (\ref{GENERICcla}) one notes obvious exchange terms between the two subsystems.

The nonlinear quantum master equation (\ref{GENERICme}) for the special case of a heat bath with dissipative bracket (\ref{heatbathdbgen}) and constant $T_{\rm e}$ has previously been given in Eq.~(5.22) of \cite{Grabert82}. This result for the special case of relaxation to equilibrium was derived by means of projection-operator techniques. The thermodynamic approach of \cite{hco199} allows us to consider arbitrary thermodynamic systems as environments and provides the fully consistent description of the mutual influence of classical environments on quantum subsystems and vice versa. A simple example of a generalization of temperature control by a heat bath would be pressure control by the environment. More interesting generalizations are obtained for anisotropic environments or when the environment itself is an open system controlled from the outside. The treatment of classical open thermodynamic systems within the geometric approach to nonequilibrium thermodynamics has been developed in \cite{hco162,hco171,hco172,hco188,Sagis10}.

\section{Solution strategies}\label{sec3}
As already elaborated, the quantum master equation (\ref{GENERICme}) is nonlinear in $\rho$ through the modified operator $\Qcommu{Q}{H}_\rho$ defined in Eq.~(\ref{Atildef}). For that reason, it is important to diagonalize the density matrix $\rho$. Instead of carrying out the diagonalization of the density matrix in every time step, we propose to write evolution equations directly for the eigenvectors $\Dket{\pi_n}$ and the eigenvalues $p_n$ in the representation (\ref{densmatrix}) of $\rho$, or for the projectors $\Pi_n = \Dket{\pi_n} \Dbra{\pi_n}$ which may sometimes be more convenient to work with. For simplicity, we assume that all eigenvalues $p_n$ are pairwise different from each other.

For a master equation of the general form
\begin{equation}\label{generalme}
    \frac{d\rho}{dt} = \frac{i}{\hbar} \Qcommu{\rho}{H} + R ,
\end{equation}
with a traceless self-adjoint operator $R$ representing the irreversible contribution to the time evolution of $\rho$, we have
\begin{equation}\label{generalPievol}
    \frac{d\Pi_n}{dt} = \frac{i}{\hbar} \Qcommu{\Pi_n}{H}
    + \sum_{m \atop m \neq n} \frac{1}{p_n-p_m}
    ( \Pi_n R \Pi_m + \Pi_m R \Pi_n ) ,
\end{equation}
for the projectors and
\begin{equation}\label{generalpevol}
    \frac{d p_n}{dt} = \Dbra{\pi_n} R \Dket{\pi_n} =
    {\rm tr} ( \Pi_n R \Pi_n ) ,
\end{equation}
for the eigenvalues. It is straightforward to verify that the separate evolution equations (\ref{generalPievol}) and (\ref{generalpevol}) imply the master equation (\ref{generalme}) by using the representation (\ref{densmatrix}) of the density matrix and the product rule.

The evolution of the eigenvectors can be expressed as
\begin{equation}\label{generalEVevol}
    \frac{d\Dket{\pi_n}}{dt} = - \frac{i}{\hbar} H \Dket{\pi_n}
    + \sum_{m \atop m \neq n} \frac{1}{p_n-p_m}
    \Dket{\pi_m} \Dbra{\pi_m} R \Dket{\pi_n}  ,
\end{equation}
which reproduces Eq.~(\ref{generalPievol}). Addition of $i X_n \Dket{\pi_n}$ to Eq.~(\ref{generalEVevol}), with an arbitrary choice of the real phase shift parameters $X_n$, would still be possible. Equation (\ref{generalEVevol}) may be considered as a modification of the Schr\"odinger equation for the eigenstates of the density matrix in the presence of dissipation caused by a perturbation from the environment. It looks very similar in structure to the result of time-independent first-order perturbation theory. Note, however, that Eq.~(\ref{generalEVevol}) describes the rate of change of $\Dket{\pi_n}$ rather than a small perturbation of $\Dket{\pi_n}$.

For the actual solution of the combined Eqs.~(\ref{generalpevol}) and (\ref{generalEVevol}), it is important to be able to evaluate matrix elements of the form $\Dbra{\pi_m} R \Dket{\pi_n}$. A comparison of Eqs.~(\ref{GENERICme}) and (\ref{generalme}) shows that $R$ contains products of the form $Q \Qcommu{Q}{H}_\rho$, $\Qcommu{Q}{H}_\rho Q$, $Q Q$, and $Q \rho Q$, which can all be evaluated after introducing a single partition of unity in terms of the eigenstates of the statistical operator. In particular, the resulting matrix elements of $\Qcommu{Q}{H}_\rho$ can then be evaluated by means of Eq.~(\ref{Arhoident}). The right-hand side of Eq.~(\ref{generalEVevol}) then contains terms of first, third, and fifth order in $\Dket{\pi_n}$ and irrational but elementary functions of $p_n$. In comparison, solution of the linearized master equation (\ref{GENERICmelin}) is considerably simpler because there is no need to diagonalize the density matrix $\rho$.

\section{Example: Two-level system}\label{sec4p}
As our first example, we consider the two-level system. In spite of its simplicity, the two-level system has successfully been used to describe both nuclear magnetic resonance and spontaneous emission in quantum optics \cite{Hahn97}.

\subsection{Notation}
For a $k$-state (or, $k$-level) system, the underlying Hilbert space is a $k$-dimensional complex vector space which, without loss of generality, we can take as $\mathbb{C}^k$. The space of observables is the $k^2$-dimensional real vector space of self-adjoint $k \times k$-matrices with complex entries. For the two-level system, we choose the $2 \times 2$-unit matrix $I$ and the three Pauli matrices
\begin{equation}\label{Pauli}
    \sigma_1 = \left(
      \begin{array}{rr}
        0 & 1 \\
        1 & 0 \\
      \end{array}
    \right) , \,\,
    \sigma_2 = \left(
      \begin{array}{rr}
        0 & -i \\
        i & 0 \\
      \end{array}
    \right) , \,\,
    \sigma_3 = \left(
      \begin{array}{rr}
        1 & 0 \\
        0 & -1 \\
      \end{array}
    \right) ,
\end{equation}
as the base vectors of the space of observables. More precisely, we express every self-adjoint complex $2 \times 2$-matrix $A$ in terms of a real scalar $\alpha$ and a real three-vector $\bm{a} = (a_1, a_2, a_3)$,
\begin{equation}\label{Arepresent}
    A = {\cal O}(\alpha, \bm{a}) = \frac{1}{2} \left( \alpha {\rm I}
    + a_1 \sigma_1 + a_2 \sigma_2 + a_3 \sigma_3 \right) .
\end{equation}
Note that $\alpha$ is the trace of $A$. Commutators and anticommutators can then conveniently be expressed as
\begin{equation}\label{commutator}
    \Qcommu{A}{B} = i {\cal O}(0,\bm{a}\times\bm{b}) ,
\end{equation}
and
\begin{equation}\label{acommutator}
    \Qantico{A}{B} = {\cal O}(\alpha\beta + \bm{a}\cdot\bm{b}, \beta\bm{a} + \alpha\bm{b} ) .
\end{equation}
From Eq.~(\ref{commutator}), we obtain an identity for the frequently occurring double commutators,
\begin{equation}\label{commutator2}
    \Qcommu{A}{\Qcommu{A}{B}} = {\cal O} (0, [a^2 \bm{1} - \bm{a}\bm{a}] \cdot \bm{b} ) ,
\end{equation}
where $a=|\bm{a}|$ and $\bm{1}$ is the $3 \times 3$-unit matrix. From Eq.~(\ref{acommutator}), we obtain
\begin{equation}\label{traceAB}
    2 \, {\rm tr}(AB) = \alpha\beta + \bm{a}\cdot\bm{b} .
\end{equation}
Arbitrary functions $f$ of an observable $A$ can be calculated with the formula
\begin{equation}\label{functionA1}
    f(A) = {\cal O} \left( f_+ + f_- , [f_+ - f_-] \, \bm{a}/a \right) ,
\end{equation}
with
\begin{equation}\label{functionA2}
    f_+ = f\left(\frac{\alpha+a}{2}\right) , \qquad
    f_- = f\left(\frac{\alpha-a}{2}\right) .
\end{equation}
Equation (\ref{functionA1}) can be verified by induction for arbitrary powers of $A$ and then be generalized by Taylor expansion. From Eq.~(\ref{functionA1}) we further conclude that $(\alpha+a)/2$ and $(\alpha-a)/2$ must be the eigenvalues of $A$.

As the density matrix has trace unity, it can be written as
\begin{equation}\label{densmatBloch}
    \rho = {\cal O}(1,\bm{m}) .
\end{equation}
For the eigenvalues to be nonnegative, we need $m = |\bm{m}| \leq 1$. This set of admissible choices of $\bm{m}$ is known as the Bloch sphere. For $m=1$, one of the two eigenvalues of $\rho$ is zero and we have a pure state. From Eq.~(\ref{traceAB}), we obtain $\ave{A} = (\alpha + \bm{a}\cdot\bm{m})/2$, which implies that the $j$th component of $\bm{m}$ is given by the average $\ave{\sigma_j}$.

\begin{figure}
\centerline{\epsfxsize=5cm \epsffile{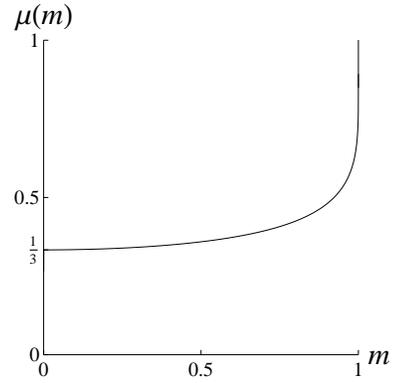}}
\caption[ ]{The function $\mu(m)$ characterizing the nonlinear contribution to the thermodynamic quantum master equation for a two-level system [see Eq.~(\ref{lambdawdef})].}
\label{fig_mu}
\end{figure}

By using Eqs.~(\ref{commutator}) and (\ref{functionA1}) in Eq.~(\ref{Atildefa2}), we find the following explicit form for the nonlinear part of $A_\rho$,
\begin{equation}\label{Atil2lev}
    A'_\rho = - \mu(m) \, {\cal O} \Big( 0,
    [ m^2 \, \bm{1} - \bm{m}\bm{m} ] \cdot \bm{a} \Big) ,
\end{equation}
with
\begin{equation}\label{lambdawdef}
    \mu(m) = \frac{1}{m^2} - \frac{1}{m \, {\rm artanh} \, m} .
\end{equation}
The function $\mu(m)$ is displayed in Figure~\ref{fig_mu}. The singularities of the two terms in Eq.~(\ref{lambdawdef}) at $m=0$ cancel so that $\mu(m) \approx 1/3$ for small $m$. According to Eq.~(\ref{commutator2}), the factor $[m^2 \, \bm{1} - \bm{m}\bm{m}]$ may be regarded as a double commutator formed with $\rho$. The nonlinear contribution to the quantum master equation associated with $\mu(m)$ leads to an improved relaxation behavior, as we shall see below.

\subsection{Bloch equation}
To arrive at an evolution equation, we now choose the Hamiltonian $H = {\cal O}(0,\hbar \omega \bm{q}_3)$, where $\omega$ is the angular frequency associated with the energy difference between the two levels of the system and $\bm{q}_3 = (0,0,1)$, as well as the two coupling operators $Q_j =  {\cal O}(0,\bm{q}_j)$ with $\bm{q}_1 = (1,0,0)$ and $\bm{q}_2 = (0,1,0)$ (note that we actually make use of the generalization mentioned at the end of Sec.~\ref{sec2B}). The environment be a heat bath, the state of which be characterized by its energy $H_{\rm e}$. According to Eq.~(\ref{tempdef}), the temperature $T_{\rm e}$ is implied by the thermodynamic relationship $S_{\rm e}(H_{\rm e})$. This temperature characterizes the black-body radiation to which our system is exposed. Both dissipative brackets are assumed to be of the form (\ref{heatbathdbgen}),
\begin{equation}\label{CalLegdb}
    \Cdissip{A_{\rm e}}{B_{\rm e}}^j = \frac{dA_{\rm e}}{dH_{\rm e}}
    \, \gamma_0 \frac{\kB T_{\rm e}}{\hbar \omega} \, \frac{dB_{\rm e}}{dH_{\rm e}} ,
\end{equation}
where $\gamma_0$ is the spontaneous emission rate.

The quantum master equation (\ref{GENERICme}) can now be recognized to be equivalent to an evolution equation for $\bm{m}$, known as the Bloch equation \cite{Bloch46},
\begin{eqnarray}
    \frac{d\bm{m}}{dt} &=& \omega \, \bm{q}_3 \times \bm{m}
    - \gamma_0 \frac{2 \kB T_{\rm e}}{\hbar \omega} \bm{R} \cdot \bm{m}
    \nonumber \\
    &-& \gamma_0 \bm{q}_3
    + \gamma_0 \frac{\mu}{2} ( m^2 \, \bm{1} + \bm{m}\bm{m} ) \cdot \bm{q}_3 ,
\label{nlBlocheq}
\end{eqnarray}
with
\begin{equation}\label{Rchoice}
    \bm{R} = \frac{1}{2} \, (\bm{1}+\bm{q}_3\bm{q}_3) .
\end{equation}
Our choice of the coupling operators $Q_j$ is motivated by the two Lindblad operators that have been derived for quantum optical applications of two-level systems for the case of spontaneous emission (see, for example, Eq.~(3.219) of \cite{BreuerPetru}). As an alternative, one could include $Q_3 =  {\cal O}(0,\bm{q}_3)$ as a third operator with the same dissipative bracket (\ref{CalLegdb}) to achieve an isotropic frictional coupling. The only effect would be to change the anisotropic matrix $\bm{R}$ in Eq.~(\ref{Rchoice}) into the unit matrix $\bm{1}$. This would correspond to the case of strong collisions that cause energy decay whenever they cause dipole phase interruption \cite{BougouffaAlAwfi08}. For nuclear spin relaxation, for which the Bloch equation had originally been proposed, the situation with $\bm{R}=\bm{1}$ can also be realized, namely in isotropic molecular environments, both in gases and in low-viscosity liquids \cite{WangsnessBloch53}. It is well known, however, that longitudinal relaxation rates that are by orders of magnitude smaller than the transverse ones are much more typical for nuclear spin relaxation \cite{Bloch46}. This situation can be achieved by enhancing the coupling strength associated with $\bm{q}_3$ to become the dominating one.

The equilibrium solution of Eq.~(\ref{nlBlocheq}) is given by
\begin{equation}\label{eqsolm}
    \bm{m}_{\rm eq} = - \bm{q}_3 \, \tanh \left( \frac{\hbar \omega}{2 \kB T_{\rm e}} \right) ,
\end{equation}
which is consistent with the canonical equilibrium distribution (\ref{rhoeq}). The occurrence of $\tanh$ in the equilibrium solution (\ref{eqsolm}) is a direct consequence of the occurrence of artanh in Eq.~(\ref{lambdawdef}). Contrary to the steady solution of the linear quantum master equation obtained from Eq.~(\ref{nlBlocheq}) for $\mu=0$,
\begin{equation}\label{nlBlocheqlin}
    \frac{d\bm{m}}{dt} = \omega \, \bm{q}_3 \times \bm{m}
    - \gamma_0 \frac{2 \kB T_{\rm e}}{\hbar \omega} \bm{R} \cdot \bm{m}
    - \gamma_0 \bm{q}_3 ,
\end{equation}
which is given by $\bm{m}_{\rm eq} = - \bm{q}_3 \hbar \omega/(2 \kB T_{\rm e})$, the steady state (\ref{eqsolm}) always lies in the Bloch sphere, even for very low temperatures. The time-dependent solution of the thermodynamic quantum master equation can actually never leave the Bloch sphere, which is a nice consequence of thermodynamic consistency. Note that, for very low temperatures, the solution of the linear master equation (\ref{nlBlocheqlin}) must leave the Bloch sphere because $m_{\rm eq} > 1$. This problem can be circumvented by the simple replacement
\begin{equation}\label{Blochrepl}
    \frac{\hbar \omega}{2 \kB T_{\rm e}} \to
    \tanh \left( \frac{\hbar \omega}{2 \kB T_{\rm e}} \right)
\end{equation}
in Eq.~(\ref{nlBlocheqlin}). With this replacement, one recovers the equilibrium result (\ref{eqsolm}) and actually obtains the well-known master equation of the Lindblad form (see Eqs.~(3.224)--(3.228) of \cite{BreuerPetru}).

For a small deviation $\bm{m}'$ from the steady state solution (\ref{eqsolm}), we obtain the following result by straightforward linearization of Eq.~(\ref{nlBlocheq}),
\begin{eqnarray}
    \frac{d\bm{m}'}{dt} &=& \omega \, \bm{q}_3 \times \bm{m}'
    - \gamma_0 \frac{2 \kB T_{\rm e}}{\hbar \omega} \bm{R} \cdot \bm{m}'
    \nonumber \\
    && \hspace{-1.5em} - \gamma_0 \frac{m_{\rm eq} \mu}{2} (\bm{1}+3\bm{q}_3\bm{q}_3) \cdot \bm{m}'
    - \gamma_0 m_{\rm eq}^2 \frac{d\mu}{d m} \, \bm{q}_3\bm{q}_3 \cdot \bm{m}' ,
    \nonumber \\ &&
\label{nlBlocheqlinx}
\end{eqnarray}
where $\mu(m)$ and its derivative are to be evaluated at $m_{\rm eq}$. The nonlinear terms enhance the relaxation in an anisotropic manner, most dramatically near the boundary of the Bloch sphere. This result of a systematic linearization of the thermodynamic master equation around the steady state is significantly different from the prediction of the usual linear Bloch equation (\ref{nlBlocheqlin}) with the replacement (\ref{Blochrepl}). The predicted strong dependence of the relaxation behavior on the location of the steady state within the Bloch sphere could be tested experimentally.

\section{Example: damped harmonic oscillator}\label{sec4}
For a detailed comparison of the linearized quantum master equation (\ref{GENERICmelin}) with the nonlinear thermodynamic master equation (\ref{GENERICme}), we study the example of the damped harmonic oscillator in one dimension. We consider the motion of a particle of mass $m$ in the potential $V(Q) = m \omega^2 Q^2 / 2$, where $\omega$ is the angular frequency of the undamped harmonic oscillator. The position and momentum are given by $Q$ and $P$ with the canonical commutation relation $\Qcommu{Q}{P}=i \hbar$, which leads us to
\begin{equation}\label{QHcomP}
    \Qcommu{Q}{H} = \frac{i\hbar}{m} P ,
\end{equation}
for $H = P^2/(2m) + V(Q)$.

For the dissipative coupling of the oscillating particle to a heat bath, we use the position $Q$ as the coupling operator in Eq.~(\ref{GENERICme}) because friction should explicitly affect only the momentum $P$ of the particle, not the position $Q$. As it is convenient to characterize friction on a particle in terms of the friction coefficient $\zeta$, we rewrite the dissipative bracket in Eq.~(\ref{heatbathdbgen}) as
\begin{equation}\label{heatbathdbCL}
    \Cdissip{A_{\rm e}}{B_{\rm e}}^Q = \frac{dA_{\rm e}}{dH_{\rm e}}
    \, \frac{\zeta \kB T_{\rm e}}{\hbar^2} \, \frac{dB_{\rm e}}{dH_{\rm e}} .
\end{equation}

It should be noted that the coupling of a harmonic oscillator to a heat bath is not unique. We here have chosen a coupling for which dissipation directly affects $P$ only. This corresponds to our intuition for the motion of a particle in a potential. However, one could alternatively assume a symmetric coupling in which both $Q$ and $P$ are explicitly affected by dissipation. This has actually been done in Eq.~(3.307) of \cite{BreuerPetru}. Such an assumption is natural, for example, for a harmonic oscillator representing an electromagnetic field mode inside a cavity.

\subsection{Caldeira-Leggett master equation}
After inserting Eqs.~(\ref{QHcomP}) and (\ref{heatbathdbCL}) into the linear quantum master equation (\ref{GENERICmelin}) for a particle moving in a potential and damped by friction with a heat bath is of the form
\begin{equation}\label{CLme}
    \frac{d\rho}{dt} = \frac{i}{\hbar} \Qcommu{\rho}{H}
    - \frac{i}{\hbar} \frac{\zeta}{2m} \, \Qcommu{Q}{\Qantico{P}{\rho}}
    - \frac{\zeta \kB T_{\rm e}}{\hbar^2} \, \Qcommu{Q}{\Qcommu{Q}{\rho}} .
\end{equation}
This equation is known as the Caldeira-Leggett master equation (see, for example, Equation (3.410) of \cite{BreuerPetru}). It is actually valid for an arbitrary potential $V(Q)$; the assumption of a harmonic potential is needed only for the next step. The Caldeira-Leggett equation cannot be brought into Lindblad form (see p.~178 of \cite{BreuerPetru}).

From second moment equations (3.428)--(3.430) of \cite{BreuerPetru}, which follow from the Caldeira-Leggett master equation (\ref{CLme}), we obtain the following closed linear differential equation for the second moment of $P$ for a harmonic oscillator,
\begin{eqnarray}
    \left[ \frac{d^3}{dt^3} + 3 \frac{\zeta}{m} \frac{d^2}{dt^2}
    + \left(4 \omega^2 + 2\frac{\zeta^2}{m^2}\right)  \frac{d}{dt}
    + 4 \omega^2 \frac{\zeta}{m} \right] \aveb{P^2} = && \nonumber\\
    4 \omega^2 \kB T_{\rm e} \zeta . \qquad &&
\label{diffeqharo}
\end{eqnarray}
If, for the initial state, the eigenvectors of the density matrix coincide with those of the Hamiltonian (for example, if we start from an equilibrium ensemble or an energy eigenstate), then the initial conditions for the differential equation (\ref{diffeqharo}) can be expressed as
\begin{eqnarray}
    \aveb{P^2} &=& \aveb{P^2}_0 , \nonumber\\
    \frac{d\aveb{P^2}}{dt} &=& - 2 \frac{\zeta}{m}
    \Big(\aveb{P^2}_0 - m \kB T_{\rm e}\Big) , \nonumber\\
    \frac{d^2\aveb{P^2}}{dt^2} &=& 4 \frac{\zeta^2}{m^2} \Big(\aveb{P^2}_0 - m \kB T_{\rm e}\Big) .
\label{diffeqharoini}
\end{eqnarray}
The explicit solution of the linear differential equation (\ref{diffeqharo}) with the initial conditions (\ref{diffeqharoini}) is given by
\begin{eqnarray}
    \aveb{P^2}_t &=& m \kB T_{\rm e} +
    \frac{\aveb{P^2}_0 - m \kB T_{\rm e}}{4\omega^2-\zeta^2/m^2} \, e^{-\zeta t/m}
    \nonumber\\
    &\times& \Bigg[ 4\omega^2 - \frac{\zeta^2}{m^2}
    \cos \left( \sqrt{4\omega^2 - \frac{\zeta^2}{m^2}} \, t \right)
    \nonumber\\
    &-& \frac{\zeta}{m} \sqrt{4\omega^2 - \frac{\zeta^2}{m^2}} \,
    \sin \left( \sqrt{4\omega^2 - \frac{\zeta^2}{m^2}} \, t \right) \Bigg] . \qquad
\label{diffeqharosol}
\end{eqnarray}

The average $\aveb{P^2}$ changes exponentially with superimposed oscillations from the  initial level to the final one, where the decay rate $\zeta/m$ in the exponential also reduces the angular frequency of the harmonic oscillator. Results for other second moments could be obtained just as easily. Equation (\ref{diffeqharosol}) contains our reference result for the damped harmonic oscillator in the usual linear description.

\subsection{Thermodynamic quantum master equation}
The thermodynamic quantum master equation for the damped harmonic oscillator is obtained by inserting Eqs.~(\ref{QHcomP}) and (\ref{heatbathdbCL}) into the quantum master equation (\ref{GENERICme}) to obtain
\begin{equation}\label{fullparticleme}
    \frac{d\rho}{dt} = \frac{i}{\hbar} \Qcommu{\rho}{H}
    - \frac{i}{\hbar} \frac{\zeta}{m} \, \Qcommu{Q}{P_\rho}
    - \frac{\zeta \kB T_{\rm e}}{\hbar^2} \, \Qcommu{Q}{\Qcommu{Q}{\rho}} .
\end{equation}
In contrast to the Caldeira-Leggett master equation (\ref{CLme}), the thermodynamic master equation is seriously nonlinear in $\rho$, even for the harmonic oscillator. For a detailed comparison, we hence need numerical solutions for concrete situations.

We consider the evolution that takes place if we start with a system equilibrized at $\kB T_0 = (3/2) \hbar \omega$ and, at $t=0$, quench the bath temperature to $\kB T_{\rm e} = (1/2) \hbar \omega$. These initial and final values of the temperature in energy units correspond to the first excited and ground state energies of the harmonic oscillator, respectively. Of course, these low temperatures have been chosen to see pronounced quantum effects. During relaxation, the probability of the ground state increases from $0.49$ to $0.86$. According to \cite{Grabert06}, the validity of the quantum master equation for such low temperatures can be established if the friction coefficient is sufficiently small. We here assume $\zeta/m=\omega/10$ to be on the safe side.

For the representation of all operators and the time-dependent eigenstates $\Dket{\pi_n}$ of the density matrix, we use a finite number of energy eigenstates $\Dket{n}$, $n=0, \ldots N$. The Hamiltonian is then represented by the diagonal matrix
\begin{equation}\label{Hmatrix}
    H = \hbar \omega \left(
          \begin{array}{cccc}
            \frac{1}{2} & 0 &   & 0 \\
            0 & \frac{3}{2} &   & 0 \\
              &   & \ddots &   \\
            0 & 0 &   & N+\frac{1}{2}
          \end{array}
        \right) ,
\end{equation}
which is exact on the truncated space. For the truncated position operator $Q$, we use the following matrix representation with nonzero entries only next to the diagonal,
\begin{equation}\label{Qmatrix}
    Q = \sqrt{\frac{\hbar}{2m\omega}} \left(
          \begin{array}{cccccc}
            0 & 1 & 0 & & & 0 \\
            1 & 0 & \sqrt{2} & 0 & & 0 \\
            0 & \sqrt{2} &  0  & \sqrt{3} & & 0 \\
            & 0 & \sqrt{3} & 0 & & 0 \\
            & & & & \ddots & \sqrt{N} \\
            0 & 0 & 0 & 0 & \sqrt{N} & 0
          \end{array}
        \right) ,
\end{equation}
which is obtained after simply omitting the couplings of the state $\Dket{N}$ to $\Dket{N+1}$. The operators $Q^2$ and $\Qcommu{Q}{H}$ are evaluated in the truncated space. For example, one obtains from Eqs.~(\ref{Hmatrix}) and (\ref{Qmatrix})
\begin{equation}\label{QHmatrix}
    \Qcommu{Q}{H} = \sqrt{\frac{\hbar^3\omega}{2m}} \left(
          \begin{array}{cccccc}
            0 & 1 & 0 & & & 0 \\
            -1 & 0 & \sqrt{2} & 0 & & 0 \\
            0 & -\sqrt{2} &  0  & \sqrt{3} & & 0 \\
            & 0 & -\sqrt{3} & 0 & & 0 \\
            & & & & \ddots & \sqrt{N} \\
            0 & 0 & 0 & 0 & -\sqrt{N} & 0
          \end{array}
        \right) .
\end{equation}
Equation (\ref{QHcomP}) can then be used to introduce the momentum operator $P$ on the truncated space in terms of this expression for $\Qcommu{Q}{H}$, and also $P^2$ is evaluated in the finite-dimensional Hilbert space.

We first implemented both the direct solution of the linear Caldeira-Leggett master equation (\ref{CLme}) for $\rho$ and the solution of the corresponding Eqs.~(\ref{generalpevol}) and (\ref{generalEVevol}) for the eigenvalues and eigenvectors of $\rho$ in Mathematica \textregistered\ (Version 7.0.1.0). As the direct solution for the linear case does not require any diagonalization, its implementation is significantly simpler and more efficient compared to solving the equations for the eigensystem given in Sec.~\ref{sec3}. Both implementations give identical results. It turns out that $10$ states ($N=9$) are sufficient to reproduce the exact results and this value is hence used for all calculations presented here. More precisely, for $10$ states, the truncation error in the initial value of $\aveb{P^2}$ is $1 \%$, which is the largest error overall because the energy and hence the importance of the higher states decreases with time.

\begin{figure}
\centerline{\epsfxsize=8cm \epsffile{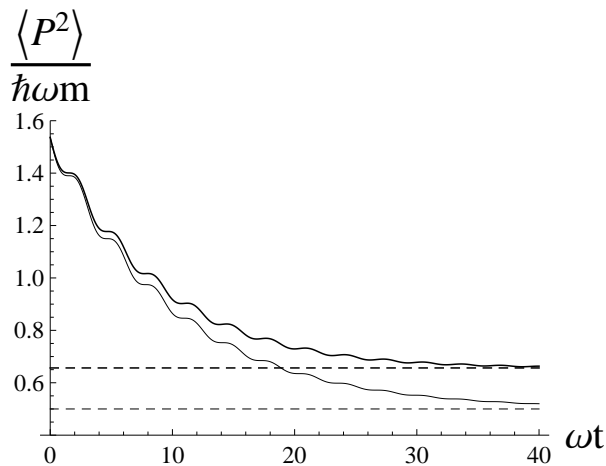}}
\caption[ ]{Relaxation of the average square of momentum after quenching a harmonic oscillator to a very low temperature for the linear (thin lines) and nonlinear (thick lines) quantum master equations. The horizontal dashed lines indicate the corresponding asymptotic values. Only the nonlinear equation leads to the correct equilibrium value (thick dashed line).}
\label{fig_psq}
\end{figure}

The evolution of $\aveb{P^2}$ and its asymptotic value are represented by the thin lines in Fig.~\ref{fig_psq}. The asymptotic value of $\aveb{P^2}$ for the linear Caldeira-Leggett master equation is given by $m\kB T_{\rm e}$, as can be seen most easily from the explicit solution (\ref{diffeqharosol}) (which is represented by the thin continuous line in Fig.~\ref{fig_psq}). This value differs from the exact value
\begin{equation}\label{P2exact}
    \aveb{P^2} = \frac{1}{2} \hbar\omega m \coth \left( \frac{\hbar\omega}{2\kB T_{\rm e}} \right) ,
\end{equation}
for a harmonic oscillator in an equilibrium state at temperature $T_{\rm e}$, which is indicated by the thick dashed line in Fig.~\ref{fig_psq}. The linearized master equation does not converge to the proper equilibrium solution at low temperatures.

For solving the nonlinear thermodynamic master equation (\ref{fullparticleme}), we us the less efficient algorithm based on the corresponding Eqs.~(\ref{generalpevol}) and (\ref{generalEVevol}) for the eigenvalues and eigenvectors of $\rho$. As only a few eigenstates are involved when we focus on low temperatures to see the nonlinear quantum effects, efficiency is is not really an issue (all calculations together took less than two minutes on a standard desktop computer). The big advantage of the algorithm based on eigensystems is that the passage from the linear to the nonlinear master equation requires a change in a single line of code only, namely the replacement described by the approximation (\ref{Arhoapprox}).

The result for the evolution of $\aveb{P^2}$ obtained from the thermodynamic quantum master equation is indicated by the thick continuous line in Fig.~\ref{fig_psq}. The solution converges to the correct limit (\ref{P2exact}) and the asymptotic value is approached faster than for the linearized equation.

\begin{figure}
\centerline{\epsfxsize=8cm \epsffile{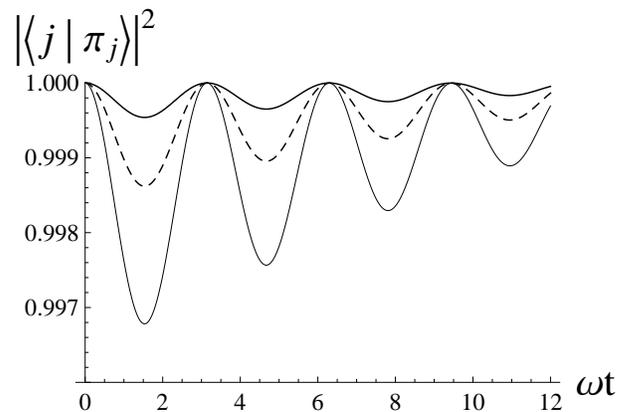}}
\caption[ ]{Overlap of the first few eigenstates of the Hamiltonian (with lowest energies) and the density matrix (with largest probabilities) after quenching a harmonic oscillator to a very low temperature as a function of time. From top to bottom, the lines correspond to the states with $j=0$, $1$, and $2$, respectively.}
\label{fig_states}
\end{figure}

The eigenstates of the initial and final equilibrium density matrices coincide with the energy eigenstates. In order to detect intermediate deviations from an equilibrium state with time-dependent temperature, we consider the matrix elements $\langle j\Dket{\pi_j}$ characterizing the overlap between corresponding eigenstates of the Hamiltonian and the density matrix. The results for $j=0$, $1$, and $2$, that is, for the states with the lowest energies and the highest probabilities, are shown in Fig.~\ref{fig_states}. The overlap turns out to be very close to unity at all intermediate times, in particular for the ground state. This is a consequence of the small value of the friction coefficient. Doubling the friction coefficient enhances the largest deviation from unity by a factor of four. For small friction, the system approximately evolves through a sequence of equilibrium states with time-dependent temperature and can hence be fully characterized by the decay of the energy, which is almost purely exponential. Note, however, that the small deviations of the overlap from unity in Fig.~\ref{fig_states} are important as they cause the oscillations of the average $\aveb{P^2}$ in Fig.~\ref{fig_psq}.

\section{Summary and conclusions}\label{sec5}
Nonlinearity is the most striking feature of the thermodynamic quantum master equation. This is a fundamental difference compared to the linear Liouville and Schr\"odinger equations describing reversible classical and quantum systems, and also to the Fokker-Planck equations for irreversible classical systems. It is the combination of irreversible thermodynamics and quantum mechanics that causes the nonlinearity. Even the master equation for the harmonic oscillator is seriously nonlinear. This fundamental nonlinearity is missed in the popular Caldeira-Leggett and Lindblad-type master equations. Just like for the linear Lindblad equations, the solutions of thermodynamic master equations stay in the physical domain for all times, which is known to be a subtle issue for nonlinear equations \cite{CzachorKuna98}. For the two-level system, the present work shows that the nonlinearity can be handled very elegantly. In general, however, the nonlinearity necessitates numerical investigations.

We have shown how the damped harmonic oscillator can be handled numerically. A coupled set of equations for the eigenvalues and eigenvectors of the density matrix is appealing because the nonlinearity can then be treated in a simple way. As the nonlinearity is a pure quantum effect, its consequences are felt only at low temperatures where only a few states are involved so that practical calculations remain feasible. Our results for the relaxation of a harmonic oscillator after a quench to a very low temperature show that the solution behavior is improved by the nonlinearity. In particular, the canonical equilibrium density matrix is approached when the nonlinearity is taken into account.

The thermodynamic quantum master equation (\ref{GENERICme}) describes the influence of any classical environment on a quantum subsystem under the assumption of weak coupling. Moreover, it is supplemented by Eq.~(\ref{GENERICcla}) describing the reverse influence of the quantum subsystem on the environment. If the total system is closed, we obtain a Markovian description of the coupled subsystems even if the coefficients in the quantum master equation change with a changing environment.

As quantum master equations are nowadays employed in many applications involving dissipative quantum systems, the nonlinear thermodynamic quantum master equation offers a new perspective on many problems. Problems that involve more complicated environments than simple heat baths can be approached in a thermodynamically consistent way. It is even possible to study situations in which the classical environment itself is an open system controlled from the outside. The treatment of classical open thermodynamic systems within the geometric approach to nonequilibrium thermodynamics has been developed in \cite{hco162,hco171,hco172,hco188,Sagis10}. If the quantum nature of the environment plays a role, one might want to consider three subsystems: the quantum system of interest, a quantum environment, and the classical environment. A coupling of either quantum system, or both quantum systems, to the classical environment would be possible.

One might ask why the thermodynamic master equation for a quantum system coupled to a heat bath proposed in \cite{Grabert82} has not been used more frequently during the past three decades. Maybe concerns about the tractability of this seriously nonlinear equation have limited its impact. The examples of the present paper should demonstrate that simple low-temperature applications can be handled quite efficiently. More complicated problems can readily be solved by well-established stochastic simulation techniques \cite{BreuerPetru,Dalibardetal92,DumZollerRitsch92,BreuerPetru95a,BreuerPetru95b} for which the nonlinearity of the thermodynamic master equation does not cause any serious difficulties \cite{hco202}.

The nonlinearity of the thermodynamic master equation for dissipative quantum systems has a number of important conceptual implications. First, as has been pointed out in the introduction and discussed in Sec.~\ref{sec2B0}, the usual ``quantum regression hypothesis'' becomes invalid and needs to be modified \cite{Grabert82,GraberTalkner83,Talkner86}. Second, it is no longer possible to pass from the Schr\"odinger picture to the Heisenberg picture because, both for reversible quantum mechanics and for linear quantum master equations, this change of pictures relies on the formal solution of the linear evolution equations in terms of exponentials (see, for example, Sec.~3.2.3 of \cite{BreuerPetru}). For dissipative quantum systems, the Schr\"odinger picture appears to be more fundamental than the Heisenberg picture. Third, as a direct consequence of the absence of a Heisenberg picture, the usual procedure for introducing two- or multi-time correlation functions fails (see, for example, Sec.~3.2.4 of \cite{BreuerPetru}). The quantum master equation describes the time evolution of the density matrix and hence also the evolution of all averages of observables for a given initial density matrix, but no two- or multi-time correlations. A natural possibility to define multi-time correlations would be to incorporate a given (positive) observable $A$ at a certain time into the density matrix $\rho$ by switching to the conditional density matrix $A_\rho$ and to continue the evolution of the master equation with $A_\rho$ instead of $\rho$. This process can be iterated several times. If the nonlinear master equation is linearized around equilibrium, the proposed procedure has actually been established to be correct for two-time correlations \cite{Grabert82}.

One of the most fascinating applications of the thermodynamic master equation, which crucially relies on its validity in the limit of low temperatures and small dissipation rates, is in quantum field theory \cite{hco200}. A friction mechanism can be used to smoothen quantum fields on short length scales by well-structured dynamic equations with well-behaved solutions. The physical origin of irreversibility in quantum field theory lies in the field idealization which requires renormalization and hence the elimination of degrees of freedom and the loss of complete control. The thermodynamic approach solves many problems of the usual renormalization program in quantum field theory which are caused by a fixed cutoff procedure that spoils the structure of the underlying reversible equations in an uncontrolled manner \cite{hco200}.

\begin{acknowledgments}
I am grateful to Hermann Grabert and Heinz-Peter Breuer for many helpful comments and suggestions.
\end{acknowledgments}



\end{document}